\documentclass[sigconf]{acmart}
\usepackage{fdsymbol}
\usepackage{xspace}
\usepackage{multicol}
\usepackage{multirow}
\usepackage{colortbl}
\usepackage{soul}
\usepackage{makecell}
\usepackage{algorithm}
\usepackage{algorithmic}
\usepackage{balance}
\usepackage{etoolbox}
\usepackage{flushend}

\AtBeginDocument{%
  \providecommand\BibTeX{{%
    \normalfont B\kern-0.5em{\scshape i\kern-0.25em b}\kern-0.8em\TeX}}}

\AtBeginDocument{%
 \providecommand\BibTeX{{%
  Bib\TeX}}}

\copyrightyear{2024}
\acmYear{2024}
\setcopyright{rightsretained}
\acmConference[KDD '24]{Proceedings of the 30th ACM SIGKDD Conference on Knowledge Discovery and Data Mining}{August 25--29, 2024}{Barcelona, Spain}
\acmBooktitle{Proceedings of the 30th ACM SIGKDD Conference on Knowledge Discovery and Data Mining (KDD '24), August 25--29, 2024, Barcelona, Spain}
\acmDOI{10.1145/3637528.3671630}
\acmISBN{979-8-4007-0490-1/24/08}

\makeatletter
\gdef\@copyrightpermission{
  \begin{minipage}{0.3\columnwidth}
   \href{https://creativecommons.org/licenses/by-nc-sa/4.0/}{\includegraphics[width=0.90\textwidth]{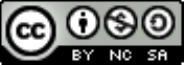}}
  \end{minipage}\hfill
  \begin{minipage}{0.7\columnwidth}
   \href{https://creativecommons.org/licenses/by-nc-sa/4.0/}{This work is licensed under a Creative Commons Attribution-NonCommercial-ShareAlike International 4.0 License.}
  \end{minipage}
  \vspace{5pt}
}
\makeatother

\begin{document}

\newcommand{\method}{{False Negative Estimation\xspace}}
\newcommand{\model}{{BHNS\xspace}}
\newcommand{\modelfull}{{Bias-mitigating Hard Negative Sampling\xspace}}
\newcommand{\sampler}{{Bias-mitigating Hard Negative Sampler\xspace}}
\newcommand\xiaochen[1]{\textcolor{purple}{\emph{#1}}}
\newcommand{\CC}[1]{\cellcolor{orange!50}}
\newcommand\haixun[1]{\textcolor{blue}{\emph{#1}}}
\title{Mitigating Pooling Bias in E-commerce Search via \\False Negative Estimation}

\author{Xiaochen Wang}
\authornote{This work was performed when the author was at Instacart.}
\affiliation{%
  \institution{The Pennsylvania State University}
  \country{University Park, PA, USA}
}
\email{xcwang@psu.edu}

\author{Xiao Xiao}
\affiliation{%
  \institution{Instacart}
 \country{San Francisco, CA, USA}
}
\email{xiao.xiao@instacart.com}

\author{Ruhan Zhang}
\affiliation{%
  \institution{Instacart}
 \country{San Francisco, CA, USA}
}
\email{ruhan.zhang@instacart.com}

\author{Xuan Zhang}
\affiliation{%
  \institution{Instacart}
 \country{San Francisco, CA, USA}
}
\email{rachel.zhang@instacart.com}

\author{Taesik Na}
\affiliation{%
  \institution{Instacart}
 \country{San Francisco, CA, USA}
}
\email{taesik.na@instacart.com}

\author{Tejaswi Tenneti}
\affiliation{%
  \institution{Instacart}
 \country{San Francisco, CA, USA}
}
\email{tejaswi.tenneti@instacart.com}

\author{Haixun Wang}
\affiliation{%
  \institution{Instacart}
 \country{San Francisco, CA, USA}
 }
 \email{haixun.wang@instacart.com}

\author{Fenglong Ma}
\affiliation{%
  \institution{The Pennsylvania State University}
 \country{University Park, PA, USA}
}
\email{fenglong@psu.edu}

\renewcommand{\shortauthors}{Xiaochen Wang et al.}

\begin{abstract}






Efficient and accurate product relevance assessment is critical for user experiences and business success. Training a proficient relevance assessment model requires high-quality query-product pairs, often obtained through negative sampling strategies. Unfortunately, current methods introduce pooling bias by mistakenly sampling
false negatives, diminishing performance and business impact. To address this, we present {\modelfull} ({\model}), a novel negative sampling strategy tailored to identify and adjust for false negatives, building upon our original {\method} algorithm. Our experiments in the Instacart search setting confirm {\model} as effective for practical e-commerce use. Furthermore, comparative analyses on public dataset showcase its domain-agnostic potential for diverse applications.
\end{abstract}





\maketitle

\section{Introduction}
\label{sec:intro}
In the rapidly evolving landscape of e-commerce search, efficient and accurate inference on query-product relevance plays a crucial role in enhancing user experiences and driving business success. 
Recently, more and more giant companies such as Instacart have used advanced techniques to enhance the performance of e-commerce searches. 
However, training such sophisticated search relevance models typically necessitates datasets of high quality and ample quantity, which is usually impractical in the e-commerce scenario due to the following reasons.  
On the one hand, annotating training data is time-consuming and labor-intensive, constraining the volume of labeled data~\cite{li2017active}. On the other hand, query-product pairs for annotation are typically pre-collected through a basic information retrieval system~\cite{cai2022hard, li2017active}. It results in a practical obstacle that labeled data are dominated by the relevant pairs, while irrelevant cases are scarce, thus harming the performance of the relevance assessment model.

To address these issues, researchers at Instacart have leveraged a transformer-based model with a negative sampling strategy to streamline the training~\cite{xie2022embedding}. Negative sampling methods facilitate the generation of irrelevant query-product pairs to complement the positive instances. This approach ensures the provision of a well-balanced training dataset that accurately mirrors the intricacies encountered in real-world e-commerce contexts. It not only aids in augmenting the model's capacity to distinguish relevant from irrelevant query-product pairs but also significantly contributes to the overall resilience and efficacy of the e-commerce search system. 

\begin{figure}[t]
\centering
\includegraphics[width=0.95\columnwidth]{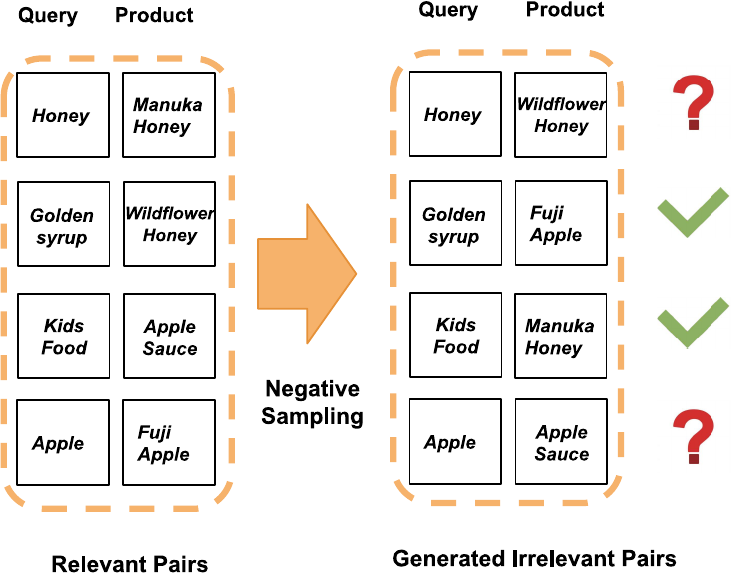}
\vspace{-0.1in}
\caption{In e-commerce scenario, conventional negative sampling usually assumes the irrelevance between original query and sampled products, which introduces pooling bias by producing false negative pairs such as [Honey, Wildflower Honey] or  [Apple, Apple Sauce]. These samples are wrongly labeled as irrelevant pairs, thus harm the performance of search model.}
\label{fig:problem}
\vspace{-0.1in}
\end{figure}

While negative sampling strategies effectively facilitate the training, they introduce another challenge: \emph{the emergence of \textbf{pooling bias} during the negative sampling process.} 
In this context, pooling bias refers to the utilization of unlabeled positive pairs as false negatives, which detrimentally impacts the model's performance~\cite{cai2022hard, arabzadeh2022shallow, thakur2021beir}. 
Conventional negative samplers assume randomly selected query-product pairs as negative (irrelevant), ignoring the possibility of false negatives. An illustration example is shown in Figure \ref{fig:problem}.

This issue becomes more pronounced for hard negative samplers that explicitly select products with high similarity to the given query, as these query-product pairs have a higher chance of being false negatives without proper control. The presence of false negatives in the training data significantly impacts the performance of the search embedding model by introducing noise, potentially leading to the overfitting issue of the model to the true positives and over-penalization of popular products. Therefore, mitigating pooling bias by appropriately addressing false negatives introduced during sampling is crucial for e-commerce search.

To tackle this problem, in this paper, we present a novel {\modelfull} ({\model}) strategy, incorporating a simple yet effective way to estimate the probability of a query-product pair being false negative, i.e., {\method}. In particular, we leverage semantic similarity between queries to calculate the false negative likelihood for query-product pairs. We then apply the estimation of false negative likelihood to eliminate pooling bias by (1) directing the sampling through regularizing the hard negative sampler and (2) guiding the training via generating pseudo labels for enhancement. We combine these two approaches and build our bias-mitigation sampling strategy. We utilize the proposed {\model} to optimize cross encoder, which takes concatenated query-product pairs as input and serves as advanced relevance assessment model \cite{rendle09bpr, he2017neural}.

To validate the efficacy of our proposed {\model}, we conduct three experiments: (1) validation with public datasets, (2) offline test with Instacart's in-house data, and (3) test as part of Instacart's search system. We compare {\model} with conventional negative sampling strategy as well as baselines specifically designed to handle false negative bias to examine the efficacy of {\model}. Through our experimental results and case study, our proposed {\model} demonstrates superior performance, thus providing strong evidence that validates the hypotheses underlying {\method} and supports the correctness of the rationale behind the design of {\model}.
\section{Background}

\begin{table}[t]
\caption{Notation table.}
    \label{tab:notation}
    \vspace{-0.1in}
    \centering
    \begin{tabular}{c|l}
    \hline
    \textbf{Notation}&\textbf{Meaning}\\\hline
     $\mathbf{e}^q_i$/$\mathbf{e}^p_j$   & the embedding of a query $q_i$/ a product $p_j$ \\ \hline
     $r_{i,j}$ & \makecell[l]{the known annotated relevance score between\\ $q_i$ and $p_j$ }\\ \hline
     $\hat{r}_{i,j}$ & the similarity score between $q_i$ and $p_j$ \\ \hline
     $\tilde{r}_{i,k}$ & \makecell[l]{the estimated relevance score between augmented\\ negative pair $q_i$ and $p_k$ } \\ \hline
     $\theta_{i,j}$ & \makecell[l]{the estimated probability of a pair $q_i$ and $p_k$ being \\as a false negative sample}  \\ \hline
     
    \end{tabular}
\end{table}

The general relevance assessment task can be formulated as follows: Each training batch is composed of two distinct components: a set of queries $\mathcal{Q}$ and a corresponding set of products $\mathcal{P}$. Given a search query $q_i \in \mathcal{Q}$ and a product  $p_j\in \mathcal{P}$, the model is responsible for assessing the degree of relevance between $q_j$ and $p_j$, which is denoted by $\hat{r}_{i,j}$. In this section, we briefly introduce the existing query-product relevance model at Instacart, empowered by conventional negative sampling strategies and more effective cross-encoder architecture in development, as the technical context of this work. Table~\ref{tab:notation} lists the key notations used in this paper.

\subsection{Search Model at Instacart}
\subsubsection{Search Embedding}\label{sec:search_embedding}
Instacart has an effective and efficient embedding model to measure the relevance between queries and products, which is widely used in search retrieval, ranking, and many other surfaces \cite{xie2022embedding}. The search embedding model follows a typical bi-encoder design, where queries and products are passed through their corresponding Sentence Transformer-based encoders\footnote{https://huggingface.co/sentence-transformers}, denoted as $\text{BE}_{Q}$ and $\text{BE}_{P}$, independently. The input query-product pair $(q_i, p_j)$ is transformed into dense embeddings as follows:
\begin{equation}
\label{eq:encoding}
\begin{gathered}
    \mathbf{e}_i^q =\text{BE}_{Q}(q_{i}),\\
    \mathbf{e}_j^p =\text{BE}_{P}(p_{j}).
    \end{gathered}
\end{equation}
The estimated relevance score between a query and a product is computed via cosine similarity between their embeddings, i.e.,
\begin{equation}
\label{eq:sim}
    \hat{{r}}_{i,j} = \text{sim}(\mathbf{e}_i^q, \mathbf{e}_j^p).
\end{equation}

During the training phase, similarity scores can be efficiently calculated through a series of matrix transpositions and multiplications, making the process computationally efficient.
For the inference stage, the product and known query embeddings can be pre-computed offline and cached, while tail query embeddings are generated in real-time.
During inference, relevant products for an incoming search query are retrieved using the query's embedding via approximate nearest neighbor search (ANN), and the estimated relevance score $\hat{{r}}_{i,j}$ is passed to the ranker service as a feature to determine the final ranking of $p_j$ for search term $q_i$. 

Although the supervised training of the search embedding model benefits from supervised learning using ground-truth annotation score $r_{i,j}$, it still suffers from the insufficiency of negative cases. In e-commerce search, positive pairs $(q_i, p_j)$ can be easily obtained from past search logs with converted products. However, high-quality negative cases, which should have been able to help the model acquire the capability of distinguishing irrelevant products from relevant ones, are naturally absent. 

\subsubsection{Negative Sampling}
A straightforward solution is selecting products that do not result in search conversions as negative examples for a given query. Nevertheless, unconverted products may not necessarily be irrelevant to the search query because users' decisions can be swayed by many factors other than relevance, such as price, size, visual presentation, personal preference for flavor or brand, item position, etc. This is confirmed in our data verification process in examining search logs from Instacart, a leading e-commerce player in online grocery shopping, where \textbf{we found more than 4\% of query-product pairs from the search logs with zero historical conversions are labeled as positive (relevant) by human annotators}. 

A common alternative approach to address the lack of negative samples issue is in-batch negative sampling. Given a query $q_i$ and in-batch products $\mathcal{P}$, let $\mathcal{P}^{+}_i$ be the set of products whose relevance with the query $q_i$ are explicitly labeled. Then the vanilla sampling process aims to randomly select $K$ products from the difference of sets $\mathcal{P}$ and $\mathcal{P}^{+}_i$, i.e., $\mathcal{P} - \mathcal{P}^{+}_i$, as negative sample set $\mathcal{P}^{-}_i$, satisfying
\begin{equation}\label{vanilla}
    \tilde{r}_{i, k} = 0, \forall p_k \in \mathcal{P}^{-}_i 
\end{equation}
where $\tilde{r}_{i,k} =0 $ denotes the irrelevance between query $q_i$ and any randomly sampled product $p_k$ in negative sample set $\mathcal{P}^{-}_i$.

Although the naive random sampling technique can generate negative pairs, it fails to utilize the most informative negative samples, resulting in insufficient learning. To avoid this issue, 
several studies~\cite{kalantidis2020hard, robinson2020contrastive} have tried to use ``hard'' negative samples during the sampling procedure due to their informativeness. At Instacart, we implement the selection of hard negative samples using our search embedding technique introduced in Section~\ref{sec:search_embedding} and determining by the similarity scores between products and queries calculated by Eq.~\eqref{eq:sim}. 
The hard negative sampling process with a negative size equal to $K$ can be represented as follows:
\begin{equation}
\label{hard}
\begin{split}
    & \tilde{r}_{i,k}=0, \forall p_k \in \mathcal{P}_i^- = argmax_{\text{top} K} \{\hat{r}_{i,j}\}, \\
    &\hat{r}_{i,j} = \text{sim}(\mathbf{e}_i^q, \mathbf{e}_j^p), \forall p_j \in \mathcal{P} - \mathcal{P}^{+}_i.
\end{split}
\end{equation}

However, indiscriminately assuming irrelevance between a given query $q_i$ and in batch product $p_k \in \mathcal{P}_i^-$ will inevitably introduce false negatives. Our case study based on the Instacart search log shows that for the query ``\textit{Honey}'', the product ``\textit{Honey 100\%}'', which is definitely relevant to the query but is not the product in this pair, will be retrieved as the top-ranked hard negative sample by the hard negative sampler mentioned in Eq.~\eqref{hard}, clearly proving the existence of pooling bias. The pooling bias diminishes the quality of augmented negative pairs, impeding the model from conducting precise relevance assessment and necessitating a concrete solution.


\subsection{Cross Encoder in E-commerce Search}

Previous work \cite{reimers2019sentence} has demonstrated the cross-encoder as a more powerful solution to capture the relationship between two input sentences compared with the bi-encoder. The cross-encoder takes the concatenation of query and product sentences as input and directly outputs the relevance score after a series of transformations. In recent exploration at Instacart, transformer-based cross-encoder\footnote{https://www.sbert.net/examples/applications/cross-encoder/}, denoted as $\text{CE}$, is adopted for relevance assessment as follows:
\begin{equation}
    \hat{{r}}_{i,j} = \text{CE}([q_i; p_j])
\end{equation}
where $[\cdot;\cdot]$ denotes the concatenation operation. 




Since the bi-encoder model encodes the query and the product separately, once the model is trained, the embedding for a given query is fixed regardless of which product it is paired with. In contrast, in the cross-encoder, the output of each intermediate layer is determined jointly by both query and product, which allows more interactions between the two and a richer context to be extracted. 

However, the cross-encoder does not output embedding of either query or product, thus presents two primary limitations. First, while training the cross-encoder, similarity score calculation via matrix multiplication, as mentioned in Eq.~\eqref{eq:sim}, is precluded due to the lack of corresponding embeddings. The score can only be obtained through aggregating queries and products in the input stage, proposing a significant computational challenge. The problem becomes particularly severe when the parameter $K$ in Eqs.~ \eqref{vanilla} and \eqref{hard} is large, constraining the scale of negative sampling. Second, instead of simply computing the cosine similarity between the cached embeddings, the cross-encoder requires the full computation of a transformer model on the fly during inference time, which results in an $O(mn)$ computational complexity that can significantly increase the latency of the system, where $m$ denotes the number of queries and $n$ is the number of products. Given this limitation, in e-commerce, $\text{CE}$ is often adopted for re-ranking only, while bi-encoders are retained for retrieval and first-stage ranking. In this work, we follow this tradition and adopt $\text{CE}$ as part of a re-ranker for a small subset of top products pre-selected by the retrieval engine and first-stage ranking.


\section{Methodology}

\begin{figure*}[t]
\centering
\includegraphics[width=0.8\textwidth]{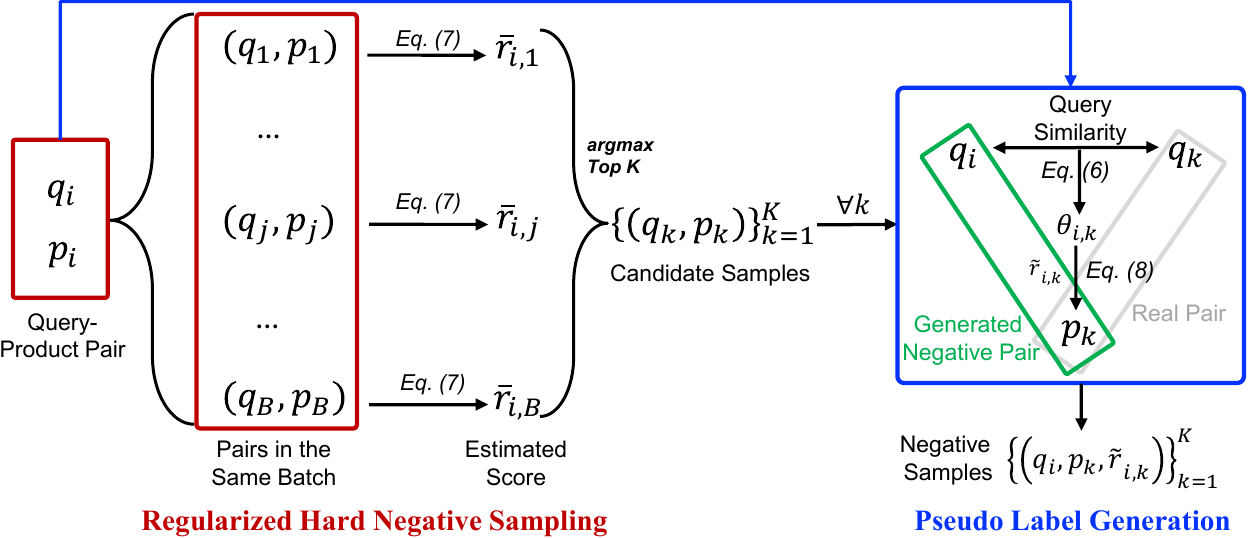}
\caption{{\modelfull}. By leveraging {\method}, the sampler detects potential false negatives and functions by (1) reducing the weight of potential false negatives during hard negative sampling and (2) informing the model of the likelihood of false negatives through the pseudo label.}
\label{fig:sampler}
\end{figure*}

To alleviate the pooling bias problem introduced above, we propose a new metric for evaluating the likelihood of false negatives, namely {\method}. Based on this metric, we design our {\modelfull} strategy, depicted in Figure \ref{fig:sampler}. Source code is available\footnote{\url{https://github.com/XiaochenWang-PSU/BHNS_Cross_Encoder}}.

\subsection{\method}
\label{sec:FNE}
Given a query $q_i$ and in-batch off-diagonal product $p_j$, our goal is to estimate the likelihood that $(q_i, p_j)$ is a false negative pair, i.e., the likelihood that $p_j$ is actually relevant to $q_i$, which is denoted as $\theta_{i,j}$. Given the fact that there is no explicit knowledge regarding the ground truth for the relationship between such in-batch pairs, we turn to estimate this relationship by leveraging Instacart's bi-encoder search embedding model mentioned in Section~\ref{sec:search_embedding} by following the hypothesis: \textit{The more similar two queries are, the more likely they share the same relevant product.}

The intuition behind this hypothesis is straightforward. Given the same target product, different users may formulate their search queries differently based on their context \cite{aula2003query}. Since the search embedding model is trained on historical search conversions, it naturally projects these different queries with similar semantic meanings to be close to each other, as well as to the product that they converted on, in the semantic space.
Similarly, if two queries are semantically independent, then it is likely that their correspondences with the same product are also mutually independent.

Based on this intuition, we do not directly estimate the relevance of a query-product pair ($q_i$, $p_j$) based on the similarity between their embeddings $\mathbf{e}^q_i$ and $\mathbf{e}^p_j$. Instead, we use the query $q_t$ that is positively related to the product $p_j$ as the bridge and take the similarity between two queries $q_i$ and $q_t$ into consideration. Since the annotation score of the pair ($q_t$, $p_j$) is known, which is $r_{t,j}$, we only need to calculate the similarity between $q_i$ and $q_t$ using Eq.~\ref{eq:sim}. Finally, we can estimate the false negative probability of a pair ($q_i$, $p_j$) as follows:
\begin{equation}\label{FNE}
    \theta_{i,j} = \frac{1}{T_j}\sum_{t=1}^{T_j} r_{t,j} * \text{sim}(\mathbf{e}^q_i, \mathbf{e}^q_t), \forall p_j \in \mathcal{P}^-_i,
\end{equation}
where $T_j$ denotes the number of queries positively related to the product $p_j$ in this batch.
The fusion of knowledge derived from the pre-trained bi-encoder and information conveyed through the ground truth annotations renders the estimation comprehensive and robust. The computation of $\text{sim}(\mathbf{e}^q_i, \mathbf{e}^q_t)$ can be performed dynamically in the data loading step without introducing significant time-consumption, as it is in the time complexity of $O(n)$.


We utilize the estimation of false negative samples in two ways for optimizing the conventional hard negative sampling process (Eq.~\eqref{hard}), i.e., sampling regularization and pseudo label generation, which will be introduced, respectively.

\subsection{Sampling Regularization}

Compared with vanilla negative sampling (Eq.~\eqref{vanilla}), the hard negative sampling strategy (Eq.~\eqref{hard}) is even more likely to introduce false negative pairs. To alleviate the problem, we propose to utilize {\method} as an additional regularization term to generate negative samples, described as follows:
\begin{equation}
\label{eq:reg}
\begin{split}
    & \tilde{r}_{i,k}=0, \forall p_k \in \mathcal{P}_i^- = argmax_{\text{top} K} \{\bar{r}_{i,j}\}, \\
    &\bar{r}_{i,j} = (1-\theta_{i,j})^\tau * \text{sim}(\mathbf{e}_i^q, \mathbf{e}_j^p), \forall p_j \in \mathcal{P} - \mathcal{P}^{+}_i,
\end{split}
\end{equation}
where $\tau$ is a temperature hyperparameter curving the weight of $\theta_{i,j}$. Here, {\method} serves as a regularization term. A larger value of $\theta_{i,j}$ implies that the pair $(q_i, p_j)$ has a high likelihood of being false negative, and Eq.~\eqref{eq:reg} ensures that the pair has less of a chance to be erroneously selected as a hard negative sample, thus balancing informativeness and robustness and mitigating pooling bias prevalent in conventional hard negative sampling. 

\subsection{Pseudo Label Generation}

Although reducing the score for potential false negatives diminishes pooling bias, unsupervised regularization cannot guarantee the extinction of all false negatives. To further eliminate the remaining pooling bias caused by these omitted cases, we design a dual insurance mechanism.
Instead of indiscriminately assuming off-diagonal query-product pairs $(q_i, p_j)$ are irrelevant as in vanilla negative sampling (Eq. \eqref{vanilla}) and hard negative sampling (Eq. \eqref{hard}), we adopt the {\method} (Eq. \eqref{FNE}) to estimate the relevance between $(q_i, p_j)$ that serves as pseudo label for training, i.e.,
\begin{equation}
\label{eq:pseudo_label}
    \tilde{r}_{i,k} = \theta_{i,k}, \forall p_k \in \mathcal{P}_i^-.
        \end{equation}

This pseudo label has the following desirable properties when $T_j=1$ in Eq.~\eqref{FNE}: 
\begin{itemize}
\item When $\text{sim}(\mathbf{e}^q_i, \mathbf{e}^q_j)$ equals 1, the two queries $q_i$ and $q_j$ are equivalent, and their relationship with product  $p_k$ is also equivalent, i.e., $\tilde{r}_{i,k} = \tilde{r}_{j,k}$. 
\item When $\text{sim}(\mathbf{e}^q_i, \mathbf{e}^q_j)$ equals 0, the two queries are independent from each other. Given that the product $p_k$ is relevant to $q_j$, it is likely irrelevant to $q_i$, and the pair $(q_i, p_k)$ is unlikely to be a false negative when we set $\tilde{r}_{i,k}$ to 0. 
\end{itemize}

In this way, the pseudo label generated via {\method} provides an extra correction on the regularized hard negative sampling strategy (Eq. \eqref{eq:reg}), further reducing the likelihood of introducing pooling bias in indiscriminate negative labeling. The combination of regularization and pseudo label based on {\method}, namely {\modelfull}, will be evaluated in the next section. Pseudo codes of the algorithm is provided in Algorithm \ref{alg:fne}.

\begin{algorithm}[]
\caption{Sampling Procedure}\label{alg:fne}
\begin{algorithmic}[1]\small
\REQUIRE Training batch $\mathrm{B}_{tra} = \{(q_0, p_0, r_{00}), \cdots, (q_B, p_B, r_{BB})$\}, frozen pretrained sentence transformer $BE_{Q}$ and $BE_{P}$.
\FOR {query $i =1$ to $B$}  
    \FOR{product $j = 1$ to $B$}
        \STATE Calculate $\bar{r}_{i,j}$ based on Eq.~\eqref{eq:reg};  
    \ENDFOR
    \STATE  Select top $K$ candidate pair set ${\{(q_k, p_k)\}_{k=0}^K}$ according to Eq.~\eqref{eq:reg};
    \FOR{product $k = 1$ to $K$}
    \STATE  Obtain $\theta_{i,k}$ according to Eq.~\eqref{FNE};
    \STATE  Calculate $\tilde{r}_{i,k}$ according to Eq.~\eqref{eq:pseudo_label};
    \STATE  Construct new pair $\{(q_i, p_k, \tilde{r}_{i,k})\}$;
    \ENDFOR
    
\ENDFOR
\end{algorithmic}
\end{algorithm}



\section{Experiment}

We conduct experiments on both public and private datasets to demonstrate the efficacy of our proposed {\modelfull} strategy. Besides, we incorporate {\sampler} with Instacart search system to validate the practical value of our proposed approach.

\subsection{Semantic Textual Similarity}

To examine the effectiveness of \method, we first conduct our experiment on the Semantic Textual Similarity (STS) task using the benchmark dataset\footnote{http://ixa2.si.ehu.eus/stswiki/index.php/STSbenchmark}. In the STS task, the similarity between two pieces of text is expected to be predicted by a model, analogous to the general search use cases.

\subsubsection{Implementation}

MiniLM-L3-v2 \footnote{https://huggingface.co/sentence-transformers/paraphrase-MiniLM-L3-v2} and MiniLM-L-2-v2 \footnote{https://huggingface.co/cross-encoder/ms-marco-MiniLM-L-2-v2} serve as bi-encoder and cross encoder in our experiments.
During the training, the bi-encoder is frozen. We set the temperature hyperparameter $\tau$ to 2.

\subsubsection{Evaluation Metric}
As the STS Benchmark evaluation dataset does not allow list-wise comparison, we decide to adopt the evaluation metrics utilized in script provided in the document of cross encoder\footnote{https://github.com/UKPLab/sentence-transformers/blob/master/examples/training/cross-encoder/training\_stsbenchmark.py}, which are Pearson correlation coefficient and Spearman correlation coefficient. Besides, we leverage the Area under the ROC Curve (AUROC) as an additional evaluation metric for a more fair comparison. These three metrics perform pair-wise evaluation simultaneously. 

\subsubsection{Baselines}
We compare our proposed {\modelfull} ({\model}) with Vanilla Negative Sampling (VNS) and Hard Negative Sampling (HNS), they serve as natural baselines widely used in diverse scenarios. Furthermore, we extend our experiment by adding the following advanced baselines specially designed for handling false negatives:
\begin{itemize}
\item \textbf{STAR} (\textbf{S}table \textbf{T}raining \textbf{A}lgorithm for
dense \textbf{R}etrieval) \cite{zhan2021optimizing} is a training pattern developed to combine random negative sampling with hard negative sampling to optimize the training process of the bi-encoder. In this study, we transfer their strategy of performing combination to the case of cross encoder.
\item \textbf{FALSE} (\textbf{F}alse neg\textbf{A}tive samp\textbf{L}es aware contra\textbf{S}tive l\textbf{E}arning model) \cite{zhang2022false} is a strategy of detecting false negative samples in an unsupervised contrastive learning scenario. We transfer this method to the supervised training of the cross-encoder by adopting their idea of capturing false negatives based on inter-sample similarity.
\item \textbf{CET} (\textbf{C}oupled \textbf{E}stimation \textbf{T}echnique) \cite{cai2022hard} leverages a subordinate cross encoder, i.e., observer,  to monitor and correct false negative bias during the training of the primary cross encoder, i.e. reranker. 
\end{itemize}

To perform a fair comparison, we also choose MiniLM-L-2-v2 as the cross-encoder for baselines. For methods involving hard negative sampling, i.e., HNS, FALSE and STAR, we adopt the bi-encoder same to BHNS (MiniLM-L3-v2) to guide the selection of hard negatives.

\begin{table}[]
\caption{Experiments on STS Benchmark Dataset. The highest score is shown in \textbf{boldface}.}
\label{STS}
\vspace{-0.1in}
\begin{tabular}{l|l|rrr}
\hline
            $K$ Valuae           &           Method                              & \multicolumn{1}{l}{Pearson} & \multicolumn{1}{l}{Spearman} & \multicolumn{1}{l}{AUROC} \\
                       \hline

\multirow{6}{*}{$K = 2$} & VNS                         & 67.61        & 77.05 & 90.18       \\
                       & HNS                     & 66.74                       & 74.57  & 89.24                    \\
                       & STAR \cite{zhan2021optimizing}                     & 59.95                       & 72.04  & 88.28                     \\
                       & FALSE \cite{zhang2022false}                         & 70.56        & 77.10 & 90.12       \\
                       & CET \cite{cai2022hard}                         & 75.87        & 75.21 & 88.96       \\
                      & \CC{}\model~(ours) & \CC{}\textbf{78.32}                       & \CC{}\textbf{77.37} & \CC{}\textbf{90.64}                       \\
                       \hline
\multirow{6}{*}{$K = 4$} & VNS                         & 67.53                       & 76.67 &  89.99                    \\
                       & HNS                    & 67.09                       & 74.11 &  88.93                      \\
                       & STAR \cite{zhan2021optimizing}                      & 62.93                       & 72.30 &  88.52               \\
                       & FALSE \cite{zhang2022false}                         & 70.76        & 76.55 & 89.76       \\
                       & CET \cite{cai2022hard}                         & 75.78        &  75.12 & 88.91       \\
                        & \CC{}{\model}~(ours) &  \textbf{\CC{}77.97}                       & \CC{} \textbf{76.91} & \CC{} \textbf{90.34}                     \\
                       \hline
\multirow{6}{*}{$K = 8$} & VNS                         & 67.49        & 76.12 & 89.90       \\
                       & HNS                  & 71.76                       & 74.81 & 89.19                      \\
                       & STAR \cite{zhan2021optimizing}                      & 65.57                      & 73.01  & 88.56            \\
                       & FALSE \cite{zhang2022false}                         & 71.85        & 76.07 & 89.43       \\
                       & CET \cite{cai2022hard}                         & 76.53        & 75.45 & 89.07       \\
                       & \CC{}\model~(ours) & \CC{}\textbf{77.30}                        & \CC{}\textbf{76.37} & \CC{}\textbf{90.05}      \\
                       \hline
\end{tabular}
\vspace{-0.1in}
\end{table}

\subsubsection{Results}
We test different sampling sizes to validate the superiority as well as the robustness of our proposed {\model}. The results are listed in Table ~\ref{STS}, where $K$ is the negative sampling size. Comparison across different sampling strategies implies that hard negative sampling may not contribute to decent performance due to the introduced false negative bias, while BHNS equipped with our proposed {\model} is enabled to handle the pooling bias, thus performing better via the bias mitigation. 

While efficiency can be regarded as a potential trouble as we introduce additional computations for bias exclusion, we observe 32\% - 38\% training time increase, compared with VNS method. Considering that inference speed will not be influenced, we confirm that our method contributes to superior performance with an acceptable compromise on efficiency.

It is also worth noting that {\model} outperforms CET, which involves an additional cross-encoder to observe and eliminate false negatives. It indicates that leveraging a frozen bi-encoder to simply estimate the likelihood of false negatives is superior to supervised sophisticated approaches demanding extra training, showcasing the soundness of the hypothesis we propose in Section \ref{sec:FNE}, as well as validating the rationale behind our design. 

\subsubsection{Ablation Study}

We also conduct ablation study on each component in {\model}. The results are presented in Table \ref{tab:ablation}. While both modules manifest competitive effectiveness compared with baselines, the pseudo label generation module performs better. Nevertheless, the combination of both modules, i.e., {\model}, is capable of achieving a more stable and superior performance. This study further validates that our design is non-redundant. 

\begin{table}[]
\caption{Ablation Study on STS Benchmark Dataset. K is set to 2 and the highest score is shown in \textbf{boldface}.}
\vspace{-0.1in}
\begin{tabular}{l|rrr}
\hline
                      & \multicolumn{1}{l}{Pearson} & \multicolumn{1}{l}{Spearman} & \multicolumn{1}{l}{AUROC} \\
                       \hline

  Pseudo Label Generation                         & 77.07        & \textbf{77.75} & 90.43       \\
                        Sampling Regulation                      & 71.57                       & 75.13  & 89.52                    \\

                       \CC{}\model~(ours) & \CC{}\textbf{78.32}                       & \CC{}77.37 & \CC{}\textbf{90.64}                       \\
                       \hline

\end{tabular}
\label{tab:ablation}
\vspace{-0.1in}
\end{table}


\subsection{Offline Experiment}\label{instacart_exp}

\subsubsection{Dataset}
\label{sec:dataset}
Datasets used for training and evaluating the cross-encoder model come from Instacart's ongoing data collection effort to benchmark search relevance, where a random sample of search queries and the products ranked in top display positions are sent for human annotation every month to judge the relevance between queries and products. The annotators categorize each (query, product) pair into five categories: 
\begin{itemize}
    \item \textbf{Strongly Relevant}: The product is exactly the type of product the search query is asking for.
    \item \textbf{Relevant}: The product fits the search query, but there are likely better alternatives.
    \item \textbf{Somewhat Relevant}: The product is not exactly what the search query asks for, but it is understandable why it shows up.
    \item \textbf{Not Relevant}: The product is not what the search query asks for, and it is unclear why it shows up.
    \item \textbf{Offensive}: The product is unacceptable and creates a bad user experience.
\end{itemize}

To help the model grasp the finer granularity in relevance, we convert the above categories into soft labels in the training dataset by taking ``Strongly Relevant'' as 1, ``Relevant'' as 0.5, ``Somewhat Relevant'' as 0.2, and ``Not Relevant/Offensive'' as 0.1. Scores from multiple annotators for the same (query, product) pair are averaged. 

Annotated classes in the evaluation dataset are converted with a coarser stroke, with ``Strongly Relevant'' and ``Relevant'' as 1, and ``Somewhat Relevant''/``Not Relevant''/``Offensive'' as 0.1, in accordance with our previous work~\cite{xie2022embedding}. 

This golden dataset with human annotation is of extremely high quality, but it can introduce bias in model training in that all samples come from products occupying top display positions in search. As a result, they are much more relevant to the search query than other products not selected in the sample. To ensure that the model is capable of correctly scoring less relevant products, we further expand both the training and the evaluation datasets by including an additional 20\% (query, product) pairs where the query and the product are randomly and independently sampled from all eligible queries and products from the full dataset. These random negative pairs get a score of 0, based on the assumption that, on average, they are even less relevant than the ``Not Relevant''/ ``Offensive'' products that appeared in top search positions.

\subsubsection{Implementation}
\label{sec:offline_implementation}
The present search embedding model serving as the bi-encoder for experiments is originally initialized with a MiniLM-L3-H384-uncased model\footnote{https://huggingface.co/nreimers/MiniLM-L6-H384-uncased} pretrained on paraphrase dataset\footnote{https://www.sbert.net/examples/training/paraphrases/README.html}. It is further finetuned using Instacart's in-house dataset extracted from the search log as described in the reference \cite{xie2022embedding}. The search embedding model is frozen during the training of the cross-encoder model and is only leveraged for guiding negative sampling. 

The cross-encoder is initialized with a pretrained MiniLM-L-2-V2 model\footnote{https://huggingface.co/cross-encoder/ms-marco-MiniLM-L-2-v2}. The chunk size of negative sampling is set to 4. The temperature hyperparameter $\tau$ is set to 2 for our method.

\subsubsection{Evaluation Metric}

We focus on the two evaluation metrics: Normalized Discounted Cumulative Gain at $M$ (NDCG@$M$), which measures the effectiveness of the model in a list-wise setting, and Area under the ROC Curve (AUROC), which measures the model's pair-wise performance. For NDCG, we aggregate the data by search queries and evaluate the model's ability to rank the products for each query by their relevance at $M$ = 5, 10, or 20. For AUROC, we convert the classes ``Strongly Relevant'' and ``Relevant'' to positive, and the classes ``Somewhat Relevant''/``Not Relevant''/``Offensive'' as well as the additional random (query, product) pairs to negative, to make the labels binary. 

\subsubsection{Baselines}

We compare our proposed {\modelfull} (BHNS) method with Vanilla Negative Sampling (VNS, Eq.~\eqref{vanilla}) and Hard Negative Sampling (HNS, Eq.~\eqref{hard}) strategies as baselines. We set the chunk size of sampled negatives to 4 for all sampling strategies implemented for fair comparisons. 

\begin{table}[t]
\caption{Offline Experiments on Instacart Search Dataset. The highest score is shown in \textbf{boldface}. ``w/o NS'' means the approach without using the negative sampling technique.}
\vspace{-0.1in}
\centering
\resizebox{1\columnwidth}{!}{
\begin{tabular}{ll|lllll}
\hline
                          & & NDCG@5          & NDCG@10         & NDCG@20         & AUROC           \\ \hline
& w/o NS      & 89.79          & 90.88          & 92.65          & 91.14          \\
& VNS                        & 90.42          & 91.43          & 93.09          & 94.09          \\
& HNS                        & 90.62          & 91.53          & 93.19          & 90.34          \\ \hline

\rowcolor{orange!50}
&\CC{}\model               &\CC{}\textbf{90.96}    &\CC{}\textbf{91.91}    &\CC{}\textbf{93.47}    &\CC{}\textbf{95.06} \\ \hline

\end{tabular}
}
\label{tab:offline}

\end{table}

\vspace{-0.1in}

\subsubsection{Results}
\label{sec: offline_results}

 We conduct our experiment as described above and obtain results listed in Table \ref{tab:offline}. The comparison showcases the advantage of adopting a false negative sampling strategy, as the cross-encoder equipped with any negative sampling methods outperforms the one without them on nearly all the metrics. Compared to VNS, HNS does not contribute to performance increase in terms of AUROC, and the performance enhancement on NCDG is minimal, implying that the pooling bias introduced during the sampling process impedes the thorough exploitation of those informative true hard negatives. As an extension, our proposed {\sampler} addresses this shortcoming of HNS by introducing \method, and thus assists cross encoder to manifest more competitive performance.

\subsubsection{Case Study}

To further demonstrate how our {\modelfull} alleviates the pooling bias in the sampling process, we perform an intuitive case study to compare {\model} with conventional negative sampling approaches, i.e., VNS and HNS. Given the query ``\textit{honey}'', we retrieve corresponding negative products using VNS, HNS, and {\model}, the results are listed in Table \ref{tab:case_study}.

It is obvious that the vanilla negative sampling method (VNS) is incapable of capturing informative hard negative samples, thus failing to boost the cross-encoder via learning from valuable sources. The hard negative sampling (HNS) strategy, although effectively collecting informative products, assumes the irrelevance between query and sampled products, thus unexpectedly generating false negatives and introducing the pooling bias. Our method, {\modelfull}, successfully avoids probable false negatives by assigning lower scores to them. Also, even though potential false negative products are collected, the pseudo label generation mechanism specially designed to tackle this scenario is able to hint the model about the potential bias, thus achieving the most promising performance.


\begin{table}[t]
    \centering
    \caption{Sampled negative products using the query ``honey''.}
\label{tab:case_study}
\vspace{-0.1in}
\resizebox{0.98\columnwidth}{!}{
    \begin{tabular}{c|l|c}
    \hline
    Approach & Product Name & Score \\\hline
     \multirow{5}{*}{VNS}    
     &bowls, compostable & 0\\\cline{2-3}
     &\makecell[l]{potato and wheat chips, cheddar, \\cheese crispers} & 0\\\cline{2-3}
     &garlic                                           & 0\\\cline{2-3}
     &classic white premium baking chips               & 0\\\hline

     \multirow{4}{*}{HNS}    
     &honey, orange blossom  & 0\\\cline{2-3}
     &peanuts, honey roasted                      & 0\\\cline{2-3}
     &moisturizing liquid hand soap milk \& honey & 0\\\cline{2-3}
     &greek strained yogurt with honey            & 0\\\hline
     
     \multirow{6}{*}{BHNS}    
     &frozen dairy dessert cookies \& cream & 0\\\cline{2-3}
     &\makecell[l]{dark chocolate, sea salt \& almonds, \\oat milk, 55\% cocoa} & 0\\\cline{2-3}
     &\makecell[l]{honey roasted breakfast cereal, whole \\grain, family size}  & 0.019\\\cline{2-3}
     &powered donuts pop'ettes                                          & 0\\\hline
     
    \end{tabular}
    }
\end{table}


\subsection{Experiment with Instacart Search System}



\begin{figure}[t]
\centering
\includegraphics[width=0.45\textwidth]{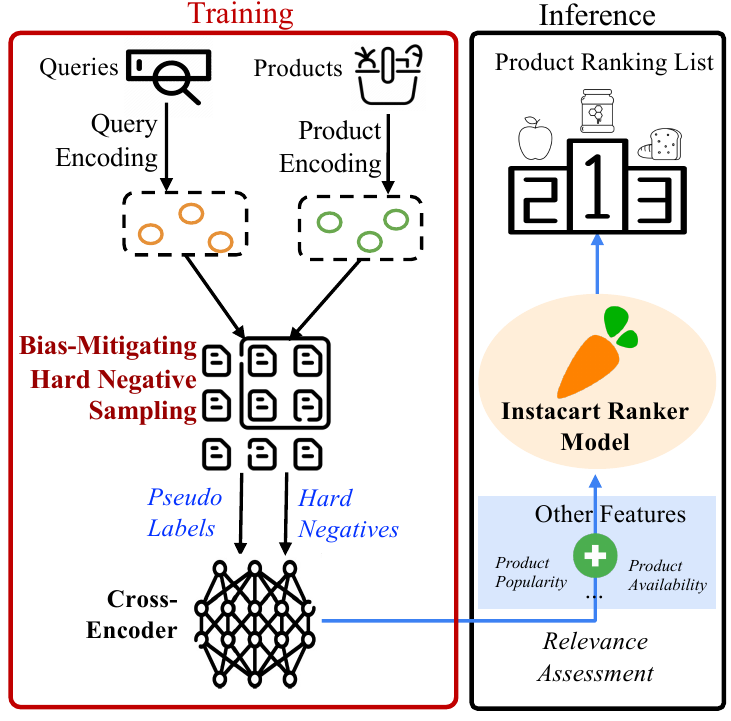}
\vspace{-0.1in}
\caption{Overview of Instacart's ranking system incorporated with \model-boosted cross encoder. The cross encoder is independently trained with Instacart search dataset (see Section \ref{sec:dataset}). During the inference time, the output of cross encoder boosted by {\model} serves as a necessary feature in the Instacart ranking system, facilitating the ranking process along with other features.}
\vspace{-0.1in}
\label{fig:system}
\end{figure}

To establish the real-world application of {\model}, we incorporate the cross-encoder trained in Section \ref{instacart_exp} with Instacart's search system and examine its impact on the relevance of search results, as shown in Figure~\ref{fig:system}.

\subsubsection{Search System at Instacart}
The search system at Instacart consists of a retrieval stage and a ranking stage. The retrieval stage casts a broad net and retrieves a large number of products related to the search query from multiple sources using both sparse and dense signals. It is followed by a multi-step ranking stage, where ranking models with increasing complexities are applied to filter out less relevant products at each step. This cascade pattern ensures that only high-quality search results are presented to the end user. Given the complexity and the latency implication of the cross-encoder model, it is tested as part of the final ranking stage, where the top 50 items get reranked to maximize search conversion. 

The final ranker is a deep \& cross network (DCN) model \cite{rxwang2017kdd} with a variety of features, such as how relevant the product is to the search query, the product's popularity and availability, the user's personal preference, etc. In the baseline ranker model serving in production, the query-product cosine similarity from the existing search embedding model \cite{xie2022embedding} is one of the dominant relevance features. In this experiment, we use the similarity score derived from the cross-encoder model, both with and without {\model}, and retrain the DCN ranker model using the same ranking objective. 
The retrained ranking is used as a replacement for the existing ranker in this test, while other components of the search pipeline remain unchanged.

\subsubsection{Evaluation Metrics}

In addition to evaluation metrics involved in Table \ref{tab:offline}, we extend to cover more evaluation metric, i.e., Mean Reciprocal Rank (MRR), for more comprehensive evaluation.

\subsubsection{Results}

We benchmark the performance of the new ranker model against the baseline model on a sample of recent searches from Instacart's business, and compare the two models' performance in ranking, where converted products are taken as positive labels. The results are listed in Table ~\ref{instacart_offline}. While the other features going into the two ranking models remain identical, the cross encoder boosted by {\modelfull} strategy proves to be effective in significantly improving both list-wise and pair-wise ranking metrics, outperforming the one without {\model}. The comparison further demonstrates the superiority of {\modelfull} in handling pooling bias in real-world e-commerce scenario.  


\begin{table}[t]
\caption{Evaluation of Instacart's Ranking Model Performance on Search Logs. The evaluation involves comparing the performance of Instacart's ranking model on search logs. In this experiment, the ranker without BHNS uses the relevance score returned by the cross encoder with a Vanilla Negative Sampling strategy. This relevance feature is replaced with the BHNS-cross-encoder score for the purpose of the experiment.}
\vspace{-0.1in}
\label{instacart_offline}
{
\begin{tabular}{l|c|c}
\hline
Metric             & Reranker w/o BHNS & Reranker w. BHNS \\ \hline
AUC          & 86.39           & \CC{}\textbf{86.46} \\
MRR          & 54.21           & \CC{}\textbf{54.31} \\
NDCG@5       & 72.19           & \CC{}\textbf{72.25} \\
NDCG@10      & 75.41           & \CC{}\textbf{75.45} \\
NDCG@20      & 76.47           & \CC{}\textbf{77.03} \\ \hline
\end{tabular}
}
\vspace{-0.1in}
\end{table}




\section{Related Work}

\subsection{Cross Encoder for Search}

Compared with a typical bi-encoder, the cross-encoder has demonstrated its more advanced capability in performing ranking~\cite{rosa2022defense}. After the emergence of BERT~\cite{devlin2018bert}, researchers take a very initial step on ranking with cross encoder by proposing monoBERT~\cite{nogueira2019passage}. Researchers further propose to utilize an advanced cross-encoder for instructing bi-encoder through knowledge distillation~\cite{choi2021improving, lin2020distilling, lin2021batch, lu2022ernie,zeng2022curriculum}. Several studies in various domains also propose to combine the bi-encoder with the cross-encoder in a cascade pattern, where the bi-encoder retriever provides informative negatives to facilitate the training of cross-encoder serving as ranking~\cite{li2021align,lei2022loopitr, zhang2021adversarial}, however, this concept does not function in our application scenario according to our internal experiments.

\subsection{Hard Negative Sampling}

Previous works have thoroughly analyzed the benefit of introducing hard negatives. Comparison between random negative sampling and hard negative sampling demonstrates the effectiveness of heuristical sampling strategy~\cite{karpukhin2020dense}. Researchers propose their method to gather global negatives via dense retrieval model~\cite{xiong2020approximate}. In the setting of contrastive learning, previous works~\cite{robinson2020contrastive, tabassum2022hard,DBLP:conf/kdd/ZhouXWNKAH21,DBLP:journals/corr/abs-2110-14844} also report that hard negatives are far more informative compared with random negatives, inducing more competitive model performance.

While the advantage of exploiting hard negatives has been acknowledged, researchers also notice the problem co-occurring with the sampling procedure. It is discerned that false negatives are introduced through the negative sampling process, thus harming the performance of models. Researchers mention that introducing hard negatives indistinguishably might result in performance degradation of bi-encoder architecture, and propose to leverage cross encoder for correcting the bias \cite{qu2020rocketqa}. Adding random negatives along with hard negatives is also proposed as an effective solution to alleviate the false negative problem~\cite{zhan2021optimizing}. A recent study on recommender system~\cite{ma2023exploring} proposes to leverage the clustering approach to eliminate false negatives and thus obtain so-called ``Real Hard Negatives''. However, the method is designed to handle sequential information and is not suitable for our pair-wise ranking scenario. 

Model leveraging similarity between positive and negative cases to estimate the probability of false negatives is also proposed in the domain of computer vision~\cite{zhang2022false}. A study in the domain of knowledge graphs also proposes that labels generated by clustering algorithms are capable of guiding models to avert false negatives~\cite{je2022entity}. Similarly, supervised models rely on data augmentation and are utilized for identifying potential false negatives~\cite{huynh2022boosting}. However, these methods require the availability of trainable embedding of entities, whose implementations are unfeasible while the cross-encoder is applied. It is also proposed that the introduction of an additional IPW-based risk model can eliminate the false negative~\cite{cai2022hard}, while its simultaneous training of two cross-encoder patterns will introduce considerable additional time consumption. 

As the conclusion of this section, most of the previous works focus on false negative problems in the setting of bi-encoder, while research on cross-encoder regarding negative sampling is rare due to the lack of independent embedding of query/product that can guide the heuristic sampling process. 
\vspace{-0.1in}
\section{Conclusion}
In this paper, we propose a novel method, referred to as {\method}, to mitigate pooling bias by identifying potential false negative samples that arise during the training of a cross-encoder for e-commerce retrieval at Instacart. {\method} utilizes query similarity to estimate the likelihood that a sampled query-product pair is a false negative. To comprehensively address this issue, we have developed two modules, namely the sampling regularization and pseudo label generation, each designed to mitigate the pooling bias introduced by false negatives. By combining these two modules, we establish the {\modelfull}, denoted as {\model}. 
To validate the effectiveness of the proposed {\model}, we conduct experiments on both public data and Instacart private data. The success of {\model} on the STS Benchmark dataset implies its applicability extends beyond the e-commerce domain.
Through a series of experiments conducted within the Instacart environment, we have demonstrated that {\model} significantly enhances the performance of the cross-encoder. This enhancement holds considerable promise for its application in the e-commerce setting. 




\vfill\eject
\begin{flushend}
\bibliographystyle{ACM-Reference-Format}
\balance
\bibliography{sample-base.bib} 
\end{flushend}


\end{document}